# Features of magnetic and magnetoelectric properties, *H-T* phase diagram of GdCr$_3$(BO$_3$)$_4$

A. Bludov[a,*], Yu. Savina[a], M. Kobets[a], V. Khrustalyov[a], V. Savitsky[a], S. Gnatchenko[a], T. Zajarniuk[b], A. Lynnyk[b], M.U. Gutowska[b], A. Szewczyk[b], N. Kuzmin[c], V. Mal'tsev[c], N. Leonyuk[c]

[a]*B. Verkin Institute for Low Temperature Physics and Engineering, NASU, 61103 Kharkov, Ukraine*
[b]*Institute of Physics, Polish Academy of Sciences, 02-668 Warsaw, Poland*
[c]*Lomonosov Moscow State University, 119991 Moscow, Russia*

[*]Corresponding author. E-mail: alexbludov1981@gmail.com, bludov@ilt.kharkov.ua

**Abstract**

Comprehensive studies of magnetic properties of GdCr$_3$(BO$_3$)$_4$ single crystal have been carried out. The integrals of intrachain ($J/k = 4 \pm 1$ K) and interchain ($J'/k \approx -0.7$ K) exchange interactions in the Cr subsystem have been determined and the strength of Cr-Gd exchange interaction ($J_{Cr-Gd}/k \sim -0.2$ K) has been estimated. The values of the exchange field $2H_E \approx 17$ T and the effective magnetic anisotropy field $H_A \approx 0.05$ T of GdCr$_3$(BO$_3$)$_4$ have been estimated. The electric polarization $\Delta P_a$ along the *a* axis in the longitudinal geometry of the experiment has been detected. Correlations between the electric polarization and the magnetization of the studied compound have been found. The spin-reorientation phase transition in the magnetically ordered state has been found. This transition exists for the external magnetic field applied along any crystallographic direction and the transition field depends weakly on the direction of the field. Nature of the spin-reorientation phase transition has been discussed. Magnetic phase diagram has been constructed and spin configurations for the low-field and high-field phases have been proposed.

**1. Introduction**

Compounds with the general formula $RM_3$(BO$_3$)$_4$ (where $R$ = Y or rare-earth elements; $M$ = Al, Ga, Fe, Cr, and Sc) attract a considerable attention over the last two decades, primarily due to their optical [1,2], magnetoelectric and magnetoelastic properties [3-5]. These borates have also high thermal and chemical stability [6]. The aluminum and iron borates are the most studied members of the family so far. The aluminum borates show good luminescent and optical properties [1,2] and demonstrate large electric polarization in external magnetic field [3]. The iron borates exhibit a variety of effects due to an interaction between the $R$ and Fe magnetic subsystems. Type of magnetic structure of the iron borates below $T_N$ (30-40 K) depends strongly on the rare-earth ion. The neodymium [7], samarium [8], europium [5], and erbium [9] iron borates are antiferromagnets with the easy-plane (EP) magnetic anisotropy. The compounds containing Pr [9], Dy [10], and Tb [11] adopt antiferromagnetic ordering with the easy-axis (EA) anisotropy. For GdFe$_3$(BO$_3$)$_4$ [12] and HoFe$_3$(BO$_3$)$_4$ [13] a spontaneous spin reorientation phase transition from the antiferromagnetic EP to EA state has been found at 9 K and 5 K, respectively. Also incommensurate magnetic phases have been discovered in GdFe$_3$(BO$_3$)$_4$ [12] and NdFe$_3$(BO$_3$)$_4$ [14]. All the investigated iron borates exhibit magnetoelectric and magnetoelastic effects, which strongly correlate with magnetic phase transitions [5,15].

At the same time, the rare-earth chromium borates $R$Cr$_3$(BO$_3$)$_4$ are much less studied members of the family though they also may exhibit diversity of magnetic, magnetoelectric, and magnetoelastic properties. Majority of them crystallizes, like other $RM_3$(BO$_3$)$_4$ compounds, in huntite-like rhombohedral structure (space group $R$32) [6], however, two high-temperature monoclinic modifications (space groups $C$2/c and $C$2) have been found too [6,16,17]. Formation of the particular structural modification depends on temperature, crystallization rate, and the rare-earth element. Coexistence of the rhombohedral ($R$32) and the monoclinic ($C$2/c) modifications in one "single" crystal [18,19] is an unwanted feature of the rare-earth chromium borates $R$Cr$_3$(BO$_3$)$_4$. However, in the case of Eu [19] and Gd [20] chromium borates, the rhombohedral modification dominates and admixture of the monoclinic one is less than 15%.

Works devoted to investigations of magnetic properties of the rare-earth chromium borates $R$Cr$_3$(BO$_3$)$_4$ are rather limited. Magnetic, thermal and optical properties of NdCr$_3$(BO$_3$)$_4$ have been reported by E.A. Popova et al. [21], where the temperature of antiferromagnetic ordering, $T_N$ = 8 K, and the values of molecular-field constants, $\lambda_{Cr-Cr}$ and $\lambda_{Cr-Nd}$, have been determined. The magnetic data point to the EP type of the antiferromagnetic order. Some experimental manifestations of low dimensionality of the magnetic structure in NdCr$_3$(BO$_3$)$_4$ have been mentioned. The value of the intrachain exchange interaction $J/k$ = 4 K has been estimated. Hereinafter, the positive sign of exchange interaction constant indicates an antiferromagnetic interaction, unless other stated. At the same time, a negative value of the Curie-Weiss temperature means, as usual, that the antiferromagnetic interaction dominates.

In the case of EuCr$_3$(BO$_3$)$_4$, the antiferromagnetic ordering appears below 9.8 K and the magnetic susceptibility shows a broad maximum at 15 K [22], which is typical of low dimensional magnets. The authors of [22] made an attempt to describe the susceptibility by using the Bonner-Fisher model of the $S$ = 1/2 chain, despite the fact that $S$ = 3/2 for Cr$^{3+}$, and determined the value of the intrachain interaction $J/k$ = 23 K. This value should be divided by the factor $S_2(S_2+1)/S_1(S_1+1) = 5$ (where $S_1 = 1/2$ and $S_2 = 3/2$) in order to estimate roughly the value of intrachain exchange interaction for the case of $S = 3/2$ chain ($J/k = 4.6$ K).

Some data on magnetic properties of the samarium chromium borate have been reported in Refs. [23,24]. For SmCr$_3$(BO$_3$)$_4$, three phase transitions were observed at temperatures $T_N = 7.8$ K, $T_2 = 6.7$ K, and $T_3 = 4.3$ K. In the single crystal studied, coexistence of the rhombohedral ($R32$) and monoclinic ($C2/c$) structures with domination of the rhombohedral one was found. The constants of the intrachain ($J/k = 2.8$ K) and interchain ($J'/k = -0.75$ K) Cr-Cr exchange interactions and of the Cr-Sm interactions ($J_{Cr-Sm}/k = -0.13$ K) have been estimated [24].

Data on the magnetic susceptibility [25] and the antiferromagnetic resonance absorption [26] of GdCr$_3$(BO$_3$)$_4$ have been published recently. The gadolinium chromium borate becomes the antiferromagnet with the EP anisotropy below $T_N \approx 7$ K and also demonstrates signatures of low dimensional magnetic properties in the paramagnetic state. The magnetic susceptibility of GdCr$_3$(BO$_3$)$_4$ above $T_N$ was described in frames of a one-dimensional model for coupled chains of the Cr$^{3+}$ ions with ferromagnetic intrachain exchange, $J/k = -6.8$ K, and antiferromagnetic interchain, $J'/k = 0.36$ K, interaction. The gadolinium subsystem has the antiferromagnetic spin-spin correlations due to the Cr-Gd exchange interaction. At the same time, there is no information on magnetic properties of GdCr$_3$(BO$_3$)$_4$ in the magnetic field higher than 1.5 T, though some suggestions of presence of a magnetic phase transition in the field range from 3 to 5 T exist [21,22]. Also, the existence of magnetoelectric effect in GdCr$_3$(BO$_3$)$_4$ may be expected, by analogy with the rare-earth iron and aluminum borates [3,5]. However, no electric polarization measurements of GdCr$_3$(BO$_3$)$_4$ have been reported till now.

The intrachain exchange interaction for compounds with Nd, Eu and Sm has the antiferromagnetic character with absolute value of the exchange constant in range from 2.8 K to 4.6 K. For GdCr$_3$(BO$_3$)$_4$, this interaction is ferromagnetic, $J/k = -6.8$ K, i.e. opposite to the common behavior. This is a rather puzzling result because the intrachain Cr-Cr distance changes only in very narrow range 3.075 - 3.095 Å for different rare-earth elements. It may point out that the model chosen in Ref. [25] is not suitable for describing magnetic properties of GdCr$_3$(BO$_3$)$_4$.

In this work, thorough studies and analysis of the magnetic, thermal and resonance properties of GdCr$_3$(BO$_3$)$_4$ single crystal are presented. The magnetic phase diagram is constructed based on experimental data on magnetization, electric polarization, and heat capacity of the crystal. To the best of our knowledge, this is the first report on electric polarization and magnetic phase diagram for the chromium $R$Cr$_3$(BO$_3$)$_4$ borates.

## 2. Experimental details

Single crystals of $GdCr_3(BO_3)_4$ were grown by spontaneous nucleation from high-temperature solution based on $K_2SO_4$-$3MoO_3$ flux. More detailed description of the synthesis can be found elsewhere [18]. The crystals were black with a greenish tint and had usual for these compounds habitus. Their sizes were about $1\times1\times1$ mm$^3$. Crystallographic axes orientations were determined by the Laue method.

Magnetization measurements were performed for the temperature range 2-300 K in magnetic field up to 5 T by means of the SQUID magnetometer MPMS-XL5 (Quantum Design). Measurements of heat capacity were carried out by using the Quantum Design PPMS system, for the temperature range 2-300 K and magnetic field up to 9 T. Magnetic resonance was studied at liquid helium temperature by using a homemade multifrequency spectrometer with a resonator (17–142 GHz, up to 7 T). Electric polarization $\Delta P(H)$ and differential magnetic susceptibility $dM/dH(H)$ were measured at 4.2 K in pulsed magnetic fields up to 15 T with a homemade setup [27,28].

## 3. Results and discussion

*3.1. Crystal structure*

The X-ray diffraction studies of both polycrystalline powder samples and single crystals of $GdCr_3(BO_3)_4$ were presented in Ref. [29]. They showed a huntite-like structure (space group $R32$) with the following unit cell parameters $a = 9.4749$ Å and $c = 7.4888$ Å. Our X-ray structural studies of the single crystals confirmed the structure mentioned above. Also, the experimental data on heat capacity and electric polarization of $GdCr_3(BO_3)_4$ presented in this work confirm that the samples have the rhombohedral structure.

It is convenient to consider the following structural elements to describe the structure of $GdCr_3(BO_3)_4$: $CrO_6$ octahedra, triangular prisms $GdO_6$ and planar triangular groups $BO_3$. The edge sharing $CrO_6$ octahedra form spiral chains parallel to the $c$ axis of the crystal. The chains are linked together by means of $GdO_6$ prisms and $BO_3$ groups into a quasi-triangular lattice. It should be noted that Cr, as well as Gd ions, are located in their own single crystallographic positions, whereas B and O ions occupy, respectively, two and three crystallographic positions. The Gd and B1 ions are located at intersections of the 3-fold and 2-fold symmetry axes. The Cr and B2 ions are located along the 2-fold symmetry axes. In figure 1, two different planar groups $B1O_3$ and $B2O_3$ are shown by yellow and brown triangles, respectively. The triangular $B1O_3$ groups are joined with $CrO_6$ chains only, while $B2O_3$, apart from being connected with three neighboring chains, are connected also with two nearest

$GdO_6$ prisms. The Cr and Gd magnetic ions are located in layers perpendicular to the *c* axis and are separated from each other by layers consisted of boron and oxygen ions. The distance between the adjacent magnetic layers is equal to ⅓ of the lattice parameter *c*. Moreover, each subsequent magnetic layer is rotated around the axis of the chain by 120 degrees. In the layers, the Cr ions form a distorted kagome lattice. The distance between the nearest Cr ions in the planes is equal to 4.827 Å, which is significantly larger than the Cr-Cr distance in the chains (3.0832 Å). The Gd-Cr distance in the layers (4.2034 Å) is larger than the one between ions located within neighboring layers (3.7414 Å).

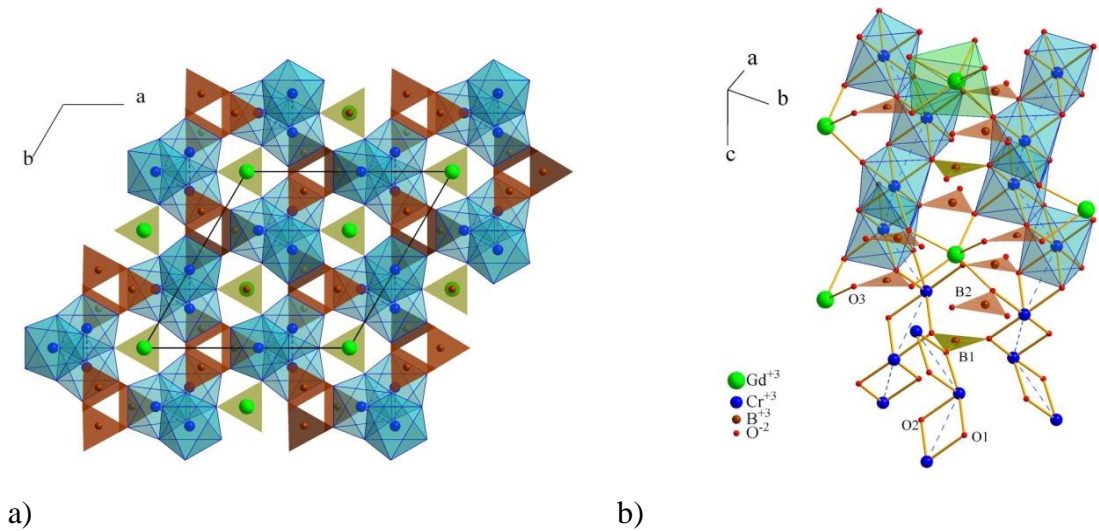

a) b)

**Fig.1.** The crystal structure of $GdCr_3(BO_3)_4$: a) view along the *c* axis, oxygen ions are omitted; b) two $Cr^{3+}$ chains and their interconnections (for details see the text).

From the point of view of magnetic properties of the studied compound, it is necessary to discuss the paths of exchange interactions in Cr subsystem (intra- and interchain ones) and exchange interactions between Cr and Gd subsystems. The intrachain exchange interaction between the nearest $Cr^{3+}$ ions realizes through a Cr-{O1,O2}-Cr planar double oxygen bridge, which has the following parameters: $d_{Cr-O1}$ = 1.9804 Å and the angle Cr-O1-Cr = 102.24°, and $d_{Cr-O2}$ = 1.9917 Å and the angle Cr-O2-Cr = 101.43°. The interchain exchange interactions between $Cr^{3+}$ ions can realize through two $B1O_3$ groups, four $B2O_3$ groups, and two $GdO_6$ prisms. So, taking into account some asymmetry of $B2O_3$, at least four different interchain exchange paths between the chains can be found. The interchain interactions connect $Cr^{3+}$ ions belonging to the same and to the adjacent layers. The $Cr^{3+}$-$Gd^{3+}$ exchange interaction path has the following parameters $d_{Cr-O3}$ = 2.1089 Å, $d_{Gd-O3}$ = 2.2196 Å, and the angle Cr-O3-Gd = 119.6° ($d_{Cr-Gd}$ = 3.7414 Å). Each Gd atom is connected with 6 Cr atoms (three of them are located in both adjacent magnetic layers). The Gd atoms are well separated from each other (the shortest $d_{Gd-Gd}$ = 6.013 Å) and exchange interaction between them can be neglected.

*3.2. Magnetic properties (T > $T_N$)*

At first, some basic information about magnetic ions in studied compound should be given. The $Gd^{3+}$ ion has half-filled $4f$ orbitals, which is a very stable $^8S_{7/2}$ configuration, with $L = 0$, $S = 7/2$ and $g_{Gd} = 1.99$. The $Cr^{3+}$ ion has $^4F_{3/2}$ ground state with $L = 3$, $S = 3/2$, and $g_{Cr} = 1.98$ [30]. The $3d$ orbitals of the $Cr^{3+}$ ion possess 3 electrons (electron configuration $d^3$). In octahedral crystal field the $d$ orbitals split into two sets of orbitals: $t_{2g}$ ($d_{xy}$, $d_{xz}$, $d_{yz}$) and $e_g$ ($d_{x^2-y^2}$, $d_{z^2}$). The $t_{2g}$ orbitals are lower in energy than the $e_g$ ones, thus, for $Cr^{3+}$, the $t_{2g}$ orbitals are half-filled and the $e_g$ ones are empty. The Goodenough-Kanamori-Anderson rules predict ferromagnetic (FM) sign of the superexchange interaction between $Cr^{3+}$ ions in octahedral environment with a common edge (the 90° exchange geometry) [31-35]. Moreover, direct overlap of $t_{2g}$ orbitals of neighbouring $Cr^{3+}$ ions gives rise to the antiferromagnetic (AFM) exchange which strongly depends on distance $d_{Cr-Cr}$ between metal atoms [31,34,35]. Total exchange interaction is a strong antiferromagnetic one for short $d_{Cr-Cr}$ and decreases very fast and then changes the sign (becomes FM) for larger $d_{Cr-Cr}$. The family of $M$CrS$_2$ ($M$ = Li, Cu, Au, Ag, Na, K) compounds [36] is a good example of such behavior of exchange interaction between $Cr^{3+}$ ions. Change of the $M$ atoms in this family leads to increase of $d_{Cr-Cr}$ and, as a result, to decrease and then to change of the sign from AFM to FM of the exchange interaction between $Cr^{3+}$ ions. Such behavior of the exchange interactions results in qualitative changes of the magnetic structure. In the case of $R$Cr$_3$(BO$_3$)$_4$ crystals with $R32$ structure, the distance $d_{Cr-Cr}$ changes only slightly with a change of the rare-earth ion and lies in the range 3.075 - 3.095 Å. It seems that for a series of chromium borates of the composition mentioned above with the rhombohedral structure, the intrachain exchange interaction varies slightly. However, the intrachain interaction ($J/k$ = -6.8 K) for GdCr$_3$(BO$_3$)$_4$ found in Ref [25], differs substantially in comparison with the data for NdCr$_3$(BO$_3$)$_4$ (4 K) [21], EuCr$_3$(BO$_3$)$_4$ (4.6 K) [22], SmCr$_3$(BO$_3$)$_4$ (2.8 K) [24], and TbCr$_3$(BO$_3$)$_4$ (3.2 K) [37]. Thus, it is important to determine the value of the exchange integral characteristic of the $Cr^{3+}$ ions in sharing edge oxygen octahedra for the range of $d_{Cr-Cr}$ mentioned above. The exchange interaction dependence on the distance between $Cr^{3+}$ ions is depicted (Fig.2) by using the published data on magnetic properties of compounds with sharing-edges CrO$_6$ octahedra [21-25,37-56] (data on organometallic compounds are not taken into account). Chemical formulae of all the compounds and numerical data presented in figure 2 are given in Appendix A. Typical value of $d_{Cr-Cr}$ for Cr ions located in edge sharing CrO$_6$ octahedra lies within the range 2.90-3.15 Å. In general, the dependence $J(d_{Cr-Cr})$ resembles qualitatively the function found for the $M$CrS$_2$ family [36]. Some dispersion of the exchange integral for different families of compounds is due mainly to structural differences. All compounds with $d_{Cr-Cr}$ < 3.098 Å have AFM exchange interaction and only NaCrGe$_2$O$_6$ ($d_{Cr-Cr}$ = 3.14 Å) demonstrates the FM one. It means that the exchange interaction changes sign at a certain $d_{Cr-Cr}$ value lying within the range 3.10 - 3.12 Å. One can conclude that the intrachain exchange interaction between Cr$^{3+}$ ions for $R$Cr$_3$(BO$_3$)$_4$ compounds with $R$32 structure should be antiferromagnetic one with the exchange integral value, expressed in the temperature unit, within the range 0.5 - 5 K. Thus, the value of the intrachain exchange interaction $J/k$ = -6.8 K for GdCr$_3$(BO$_3$)$_4$ obtained in Ref. [25] is doubtful and a more detailed analysis of the magnetic properties of this compound, taking into account new information about sign and value of the intrachain exchange interaction, must be carried out.

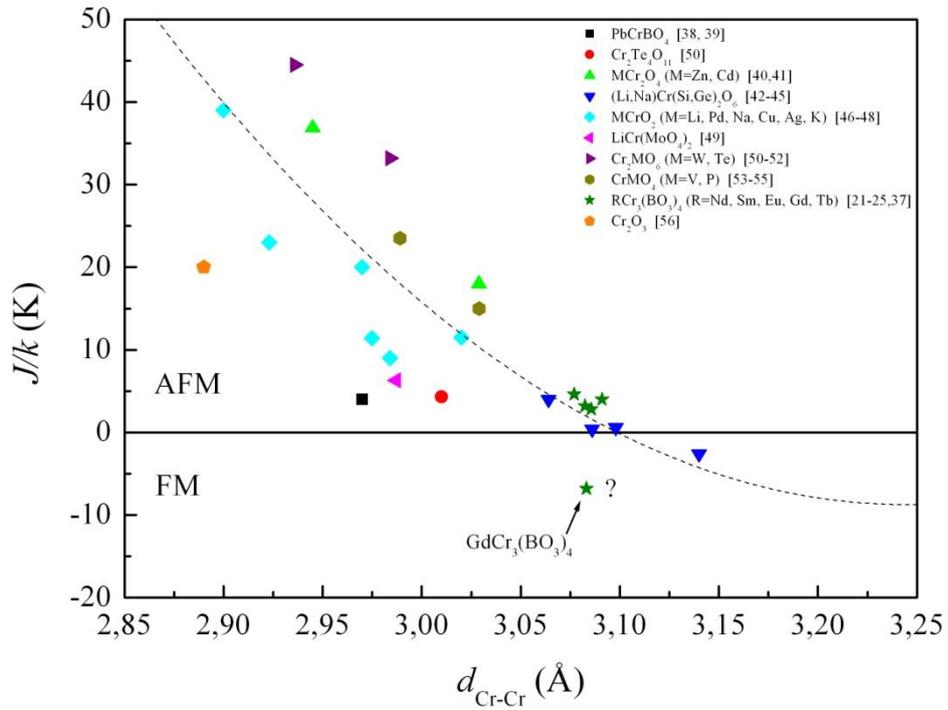

**Fig.2.** The Cr$^{3+}$-Cr$^{3+}$ exchange interaction as function of the distance between Cr$^{3+}$ ions in the octahedral oxygen environments with a common edge. Shape and color of the symbols display different groups of compounds. Dashed line shows qualitatively behavior of the $J(d_{Cr-Cr})$ dependence (drawn as a guide for eye).

Following Hamiltonian can be used to describe magnetic properties of GdCr$_3$(BO$_3$)$_4$

$$H = 2J \sum S_{Cr,i} \cdot S_{Cr,j} + 2J' \sum S_{Cr,i} \cdot S_{Cr,k} + 2J_{Cr-Gd} \sum S_{Cr,i} \cdot S_{Gd,l}. \qquad (1)$$

Here the first and the second term take into account the intra- and the interchain exchange interactions in the chromium subsystem and the third one – the exchange interactions between the chromium and the gadolinium subsystems. We assume that the Gd-Gd exchange interaction is negligibly small and excluded it from consideration. In Ref. [25] the magnetic susceptibility of GdCr$_3$(BO$_3$)$_4$ was analysed by the expression

$$\chi(T) = \frac{3\chi_{Cr}(T)}{1+\frac{2zJ'\chi_{Cr}(T)}{N_A\mu_B^2 g_{Cr}^2}} + \frac{N_A\mu_B^2 g_{Gd}^2 S_{Gd}(S_{Gd}+1)}{3k(T-\theta_{Gd})}, \qquad (2)$$

where the first and the second term describe susceptibility of the chromium and the gadolinium subsystems, respectively. It is clearly seen that the second term represents the Curie-Weiss law. In expression (2), $\chi_{Cr}(T)$ is susceptibility of a spin chain with isotropic exchange interaction, derived by Fisher for the classical limit ($S\to\infty$) [57]:

$$\chi_{Cr}(T) = \frac{N_A\mu_B^2 g_{Cr}^2 S_{Cr}(S_{Cr}+1)}{3kT} \cdot \frac{1+u(K)}{1-u(K)},$$

$$u(K) = \coth(K) - \frac{1}{K}, \qquad (3)$$

$$K = -\frac{2JS_{Cr}(S_{Cr}+1)}{3kT}.$$

In expressions (2) and (3), the following parameters are used: $J$ and $J'$ are the intra- and interchain Cr-Cr exchange interactions, $z$ is the number of Cr neighbors taken into account for the interchain interaction, by means of the parameter $\theta_{Gd}$ we try to take into account the Cr-Gd interaction (formally it is the Curie-Weiss temperature), $S_{Cr} = 3/2$ and $S_{Gd} = 7/2$ are the spins of Cr$^{3+}$ and Gd$^{3+}$ ions, $g_{Cr} \approx g_{Gd} \approx g = 2.00$ is the spin only $g$-factor, $N_A$ is the Avogadro constant, $k$ is the Boltzmann constant, $\mu_B$ is the Bohr magneton. In the previous work [25], the set of parameters for GdCr$_3$(BO$_3$)$_4$ ($J/k = -6.8$ K, $J'/k = +0.36$ K for $z = 6$ and $\theta_{Gd} = -2.8$ K) was found by fitting (2) with (3) to experimental data. This result is shown in Fig.3 by the solid red line. In this work, according to the conclusion on the magnitude and sign of the intrachain exchange, we limited the variation range of $J/k$ to 0-5 K. Then, the another set of parameters was obtained: $J/k = +4.3$ K, $J'/k = -2.8$ K for $z = 6$ and $\theta_{Gd} = +2.7$ K (dashed blue line in Fig.3). Above 10 K, both curves describe experimental data satisfactorily. Below 10 K, the difference between blue dashed line and experimental data grows abruptly due to positive value of $\theta_{Gd}$. Before further discussion of the fitting results, limitations of the approach and the following issues must be considered. Can the expression (3), obtained for the classical limit ($S\to\infty$), be used for the system with $S = 3/2$? Can the second term of the expression (2) be used for taking into account the Cr-Gd exchange interaction in the case, when the following inequality $|J| \gg |J_{Cr-Gd}|$ does not hold? The Fisher expression (3) was successfully used for analyzing magnetic properties of the chain magnets with $S = 5/2$ and $3/2$ [21,43,58,59]. This expression can be applied to both types of chains, i.e., with antiferromagnetic and ferromagnetic intrachain exchange interactions. It seems that Cr contribution to the magnetic susceptibility of $GdCr_3(BO_3)_4$ can be described by the first term of (2) with $\chi_{Cr}(T)$ given by the expression (3). The second term in expression (2), which takes into account Gd contribution requires more caution in its application. It should work well in case when $|J_{Cr-Gd}|\to 0$ ($\theta_{Gd}\to 0$) and the Curie-Weiss law transforms into the Curie law for the Gd subsystem. When the exchange between the chromium and gadolinium subsystems is finite and comparable in strength to the exchange in the chromium subsystem, the second term of (2) can be the source of some discrepancies or artefacts especially when $J$ prefers the antiferromagnetic order and $J_{Cr-Gd}$ - the ferromagnetic one or vice versa. Inability of determining the value of the $J_{Cr-Gd}$ exchange constant by using the obtained $\theta_{Gd}$ is another disadvantage of this approach. However, the sign and order of magnitude of $J_{Cr-Gd}$ can be derived by means of this simple approach.

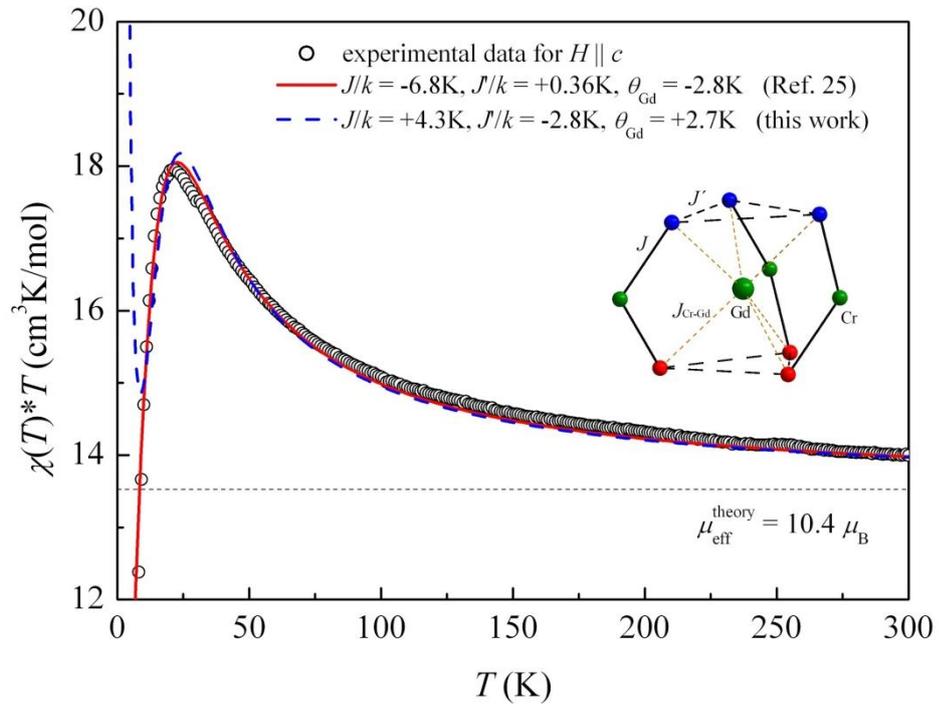

**Fig.3.** The product $\chi(T)*T$ as a function of temperature for $GdCr_3(BO_3)_4$. Symbols are experimental data for $H \parallel c$. Lines are the results of fitting of (2) to the experimental data. Inset shows the exchange interaction paths in the crystal (big sphere - $Gd^{3+}$, small spheres - $Cr^{3+}$, each colour of spheres labels magnetic ions belonging to the same magnetic layer).

Now, the obtained set of parameters ($J/k$ = +4.3 K, $J'/k$ = - 2.8 K for $z$ = 6 and $\theta_{Gd}$ = +2.7 K) will be discussed. It has been already mentioned that limitations on variation of the strength of intrachain exchange interaction were imposed a priori and the obtained value $J/k$ = +4.3 K is very close to the results obtained for compounds with Nd, Eu, Sm and Tb (see Fig.2 and Ref. [21-23,37]). So, there is not much to discuss about this parameter further. The interchain exchange interaction ($J'/k$ = - 2.8 K) has the same sign (FM) as the ones got for SmCr$_3$(BO$_3$)$_4$ (-0.75 K) [24] and TbCr$_3$(BO$_3$)$_4$ (-0.68 K) [37] but its magnitude is four times larger. Modulus of the interchain interaction constant can be estimated by using the expression obtained by Schulz [60],

$$|2J'_s| = \frac{T_N}{1.28z\sqrt{\ln(\frac{5.8*2J}{T_N})}}, \qquad (4)$$

where $T_N$ denotes the Neel temperature, $J$ - intrachain interaction constant, $J_s'$ - interchain interaction constant, $z$ - coordination number for interchain interaction. Since all parameters appearing in (4) are known, the value of the interchain interaction constant can be estimated $|J_s'/k|$ = 0.33 K. The $|J_s'/k|$ value estimated for GdCr$_3$(BO$_3$)$_4$ by using (4) and the values obtained for samarium and terbium borates are of the same order. It should be noted here that the case of GdCr$_3$(BO$_3$)$_4$ differs from the case considered by Schultz by the presence of Gd magnetic subsystem, which is coupled to the Cr chains ($J_{Cr-Gd}$) and the value $|J_s'|$ should be a certain combination of the real interchain interaction $J'$ and the Cr-Gd interaction constant $J_{Cr-Gd}$. Thus, one can conclude that the value $J'/k$ = - 2.8 K is overrated due to the limitation of the used approach. It seems that a more plausible value of interchain exchange interaction in case of GdCr$_3$(BO$_3$)$_4$ should be closer to the value of this parameter found for the counterpart compounds with samarium and terbium, $J'/k \approx$ -0.7 K. In the case when $J_{Cr-Gd}$ corresponds to the ferromagnetic coupling and it is large enough that the condition $|J| \gg |J_{Cr-Gd}|$ does not hold, the expression (2) can be applied only for temperatures much higher than $T_N$, where there are no antiferromagnetic short-range spin-spin correlations in the Cr subsystem. This is so, because these correlations affect the Gd subsystem due to the presence of $J_{Cr-Gd}$ and then the second term in (2) does not work well. Each Gd$^{3+}$ ion is connected with 6 Cr$^{3+}$ ions (three in each of two adjacent magnetic layers) and the Cr-Gd exchange interaction, for some extent, contributes to the determined $\theta_{Gd}$ and $J'/k$ values. The sign of $\theta_{Gd}$ points out to ferromagnetic character of the Cr-Gd interaction. This result can be approved by considering data on iron borates. Susceptibility of chromium and iron borates obeys the Curie-Weiss law in high-temperature range with effective magnetic parameters (effective magnetic moment - $\mu_{eff}$ and Curie-Weiss temperature - $\theta_{CW}$). The Curie-Weiss temperature is determined by contributions of all exchange interactions: $J$ - intrachain, $J'$ - interchain and $J_{M-R}$ - exchange interaction between 3$d$ transition ions (M = Cr, Fe) and rare-earth ions. One can suppose that the all mentioned interactions do not change sufficiently with changing the rare-earth ion and, in this case, only contribution from $J_{M-R}$ into $\theta_{CW}$ varies by virtue of the change of the total angular momentum $j$. The Curie-Weiss temperatures $\theta_{CW}$ for the rare-earth iron (black squares) and chromium (red circles) borates are shown in Fig.4 (data from Ref. [61]). The dotted line depicts the $j(j+1)$ value for each rare-earth ion. As one can see from figure 4, the Curie-Weiss temperature is negative for the rare-earth iron borates (varying from -51 K for $GdFe_3(BO_3)_4$ to -170 K for $EuFe_3(BO_3)_4$), which indicates the predominance of antiferromagnetic exchange interactions due to the large intrachain exchange interaction. One can also notice that, in general, there is a correlation between $\theta_{CW}$ and $j(j+1)$ (except compounds containing La and Dy). The $|\theta_{CW}|$ value decreases, when $j(j+1)$ grows. Such a behavior of $\theta_{CW}$ points out that the Fe-R exchange interaction is ferromagnetic. There are still not enough data for chromium borates but the same trend as for iron borates can be noticed for $\theta_{CW}$. Thus, one can assume that $J_{Cr-R}$ is also ferromagnetic. Another argument in favor of the ferromagnetic character of Cr-Gd interaction can be found in work published by Singh et al. [62], where the authors studied angular dependence of the exchange interaction in fluoride-bridged $Cr^{3+}$-$Gd^{3+}$ complexes and showed that the Cr-Gd exchange interaction is ferromagnetic, when the angle Cr-F-Gd is less than 140°. The magnitude of this interaction varies from 0 K to several tenths of kelvin, when the angle decreases from 140°. In our case the angle Cr-O3-Gd = 119.6° that corresponds to Cr-Gd exchange interaction from the range -0.15 - - 0.35 K. One can suppose that replacement of $F^-$ with $O^{2-}$ in the exchange bridge does not affect situation qualitatively.

The values of $\theta_{CW}$ for the gadolinium iron and chromium borates may point out that the value $J_{M-R}$ is larger in comparison with values for compounds containing other rare-earth ions. This behavior of $3d$-$4f$ exchange interaction may be attributed to the mechanism proposed by Kahn [63], based on electron transfer from a singly-occupied $3d$(Cr) orbital to an empty $5d$(Gd) orbital. It seems that this charge transfer is most probable for the $Gd^{3+}$ ions among the rare-earth elements [64].

Analysis of the published data on the $Cr^{3+}$-$Cr^{3+}$ and $Cr^{3+}$-$Gd^{3+}$ exchange interactions and the magnetic susceptibility of $GdCr_3(BO_3)_4$ gives the following values and signs of exchange interaction constants in the studied compound: the intrachain Cr-Cr interaction is antiferromagnetic $J/k = 4 \pm 1$ K, the interchain Cr-Cr interaction is ferromagnetic with $J'/k \approx -0.7$ K, and the Cr-Gd exchange interaction is also ferromagnetic with $J_{Cr-Gd}/k \sim -0.2$ K. It seems that the used approach (expression (2) and (3)) may predict the signs and order of exchange interaction values in $GdCr_3(BO_3)_4$ when applied for temperature range $T \gg T_N$. However, strictly speaking, it cannot be applied for describing the magnetic susceptibility of $GdCr_3(BO_3)_4$ due to comparatively high value of "ferromagnetic" $J_{Cr-Gd}$ and relatively high value of the effective magnetic moment, $[g^2j(j+1)]^{1/2}$, of $Gd^{3+}$ ions. We think that this approach can be valid to some extent for $RCr_3(BO_3)_4$ with light rare-earth elements ($R$ = La-Eu), which have low values of $g$-factor and total angular momentum $j$.

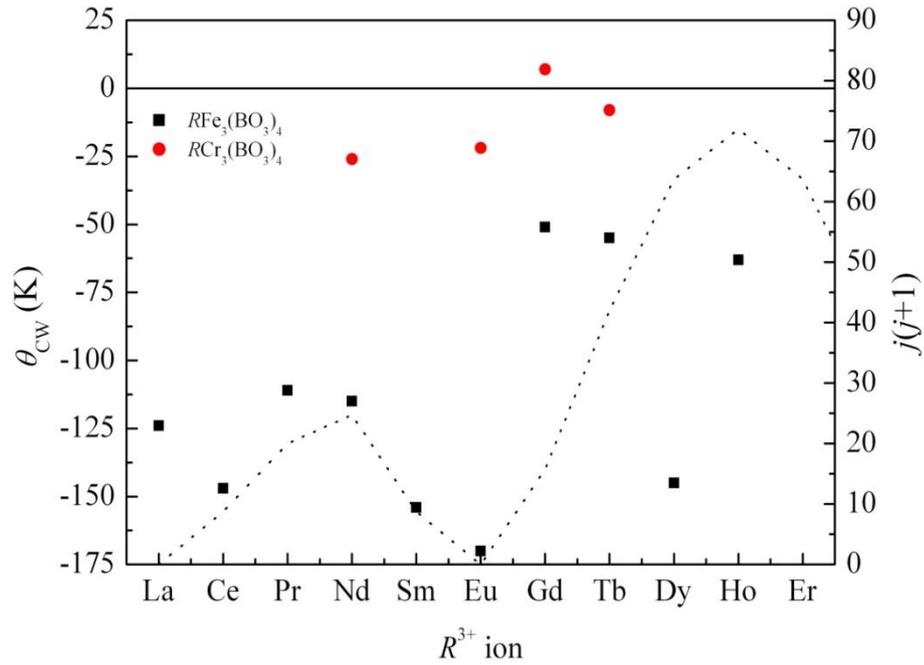

**Fig.4.** The Curie-Weiss temperature as a function of rare-earth ion for iron (black squares) and chromium (red circles) borates. Dotted curve shows the $j(j+1)$ value for each rare-earth ion (right vertical scale), where $j$ is the total angular momentum of $R^{3+}$.

*3.3. Magnetic susceptibility at low temperatures*

Figure 5 shows a low temperature part (2-15 K) of the magnetic susceptibility $\chi(T)$ of GdCr$_3$(BO$_3$)$_4$ measured in the magnetic field of $1\cdot 10^{-3}$, 0.1, 2.75, and 4.5 T, applied perpendicular to the $c$ axis of the crystal. The susceptibility curves measured in weak magnetic field ($1\cdot 10^{-3}$ and 0.1 T) coincide well in the paramagnetic region. Then, as temperature is decreasing, a sharp drop of the susceptibility is observed at the temperature of magnetic ordering, which is close to 7.1 K for both curves. The susceptibility grows slightly with further temperature declining. The temperature dependence of the magnetic susceptibility obtained for magnetic field 2.75 T is qualitatively similar to the low-field dependencies, but there is no quantitative agreement even in the paramagnetic region. The susceptibility feature which accompanies the magnetic ordering diminishes and shifts to the lower temperatures (5.8 K) with increase of the magnetic field. The curves do not coincide in the paramagnetic temperature range (above 10 K), because the susceptibility depends also on magnetic field, not only on temperature. The arrows point at the features on the susceptibility appearing at the ordering temperature, $T_N$, for each field value. The features for 0.1 and 2.75 T look like a sharp drop of the susceptibility at 7.0 and 5.8 K, respectively. Meanwhile, the feature on $\chi(T)$ measured in external magnetic field 4.5 T appears as a small growth of susceptibility with decreasing temperature below 5.1 K. The qualitative difference between the features of $\chi(T)$ associated with magnetic ordering for the magnetic field 2.75 T and 4.5 T can be a sign of a magnetic phase transition, which happens at certain field in the range between the two mentioned field values. A strong dependence of the magnetic ordering temperature on magnetic field should be noted.

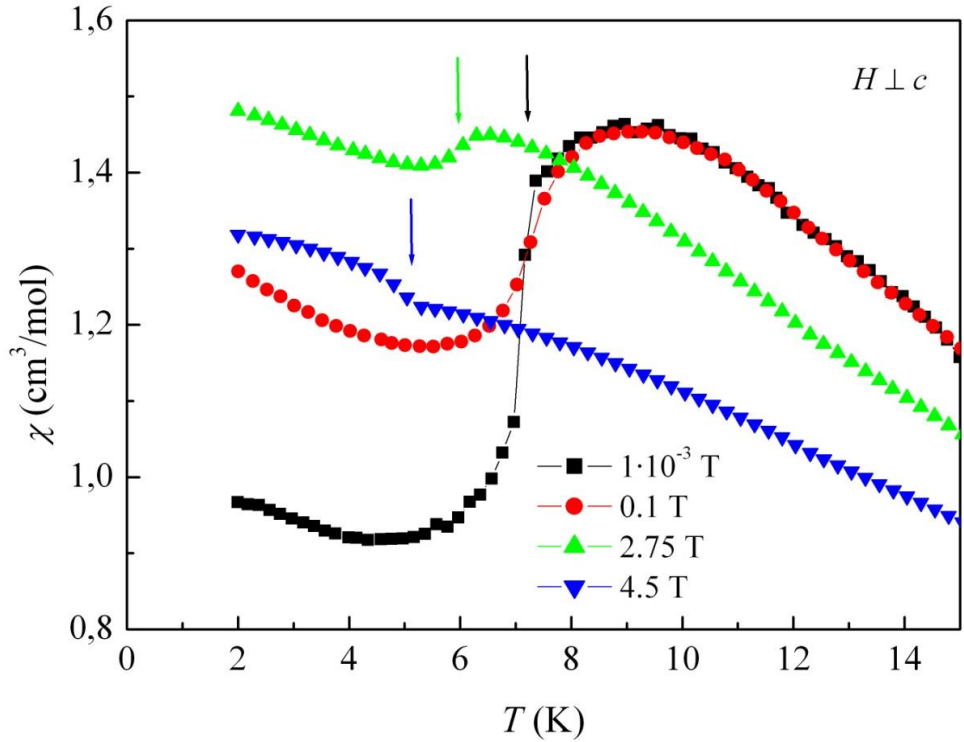

**Fig.5.** Temperature dependencies of magnetic susceptibility $\chi(T)$ of GdCr$_3$(BO$_3$)$_4$ measured in magnetic field $1 \cdot 10^{-3}$ T (black squares), 0.1 T (red circles), 2.75 T (green triangles) and 4.5 T (blue reversed triangles). Magnetic field was applied perpendicularly to the $c$ axis. Arrows point at temperatures of the magnetic ordering for each field value.

*3.4. Isothermal magnetization at $T < T_N$*

Isothermal magnetization curves $M(H)$ for GdCr$_3$(BO$_3$)$_4$ measured at 2, 4, 5 and 6 K for external magnetic field orientation $H \perp c$ are shown in Fig.6(a). The dependencies $M(H)$ for all mentioned temperatures are nonlinear in the investigated range of magnetic fields. Such a behavior is most likely due to the gadolinium subsystem contribution to magnetization. The dependencies measured at 2 and 4 K show jump of magnetization at 4.2 T, which is a sign of a spin-reorientation magnetic phase transition. This is in agreement with the results of the studies of susceptibility. The exchange field $2H_E$ of GdCr$_3$(BO$_3$)$_4$ can be estimated by extrapolating the part of $M(H)$ measured at 2 K above the phase transition with the linear function up to $M_{sat} = 16\ \mu_B$ despite its slight nonlinearity ($2H_E \approx 17$ T). No clear evidences of any features on the curves measured at 5 and 6 K are seen in Fig.6(a). Derivatives of all four

$M(H)$ curves are depicted in Fig.6(b) for more thorough analysis. It turns out that for all derivatives there is a feature in the form of a broad maximum at field close to 0.5 T. It can be attributed to transformation of the antiferromagnetic domain structure in external magnetic field. Apart from the feature at low field and the feature attributed to the spin-reorientation transition there are no additional peculiarities on curves measured at 2 and 4 K. There are two additional features at 3.5 and 4.7 T (green arrows in Fig.6(b)) on the $dM/dH(H)$ curve measured at 5 K and only one at 2.8 T (blue arrow in Fig.7(b)) for the curve measured at 6 K.

The spin-reorientation transition exists up to a certain critical temperature (4 K < $T_{cr}$ < 5 K) and its field ($H_{sr}$) does not depend on the temperature. Moreover, this transition exists for all directions of external magnetic field in the crystal and the field $H_{sr}$ varies in narrow range from 3.9 to 4.2 T (see Fig.7). Small hysteresis (curve for $\varphi = 90°$ in Fig.7) proves the first order nature of this transition. The isothermal magnetization curves are isotropic for $H < H_{sr}$ and small differences between the curves registered for different orientations of the magnetic field can be noticed above the phase transition field. We think that the presence of the spin-reorientation transition is related to the magnetic anisotropy of the $Cr^{3+}$ ions, located inside the $CrO_6$ complexes, i.e. in the octahedral crystalline electric field (Fig.7). Kinks on isothermal magnetization curve measured at 2 K were observed also for $NdCr_3(BO_3)_4$ [21] and $EuCr_3(BO_3)_4$ [22] at about 3 and 4.7 T, respectively. One can assume that the transition at about 3-5 T is a common feature for all chromium borates with $R32$ structure. However, it manifests itself differently (as a jump or a kink on the magnetization curve) for compounds with different rare-earth ions and for different orientations of the magnetic field. Pronounced nonlinearity of $M(H)$ for $GdCr_3(BO_3)_4$ at 0.5 T < $H$ < $H_{sr}$ is caused by contribution of Gd subsystem. Unlike gadolinium chromium borate, compounds containing Nd and Eu show linear behavior of isothermal magnetization in ordered state up to 3 T.

Results of magnetization measurements in pulse field (up to 15 T) for non-oriented sample at 4.2 K are shown by blue and red curves in Fig.8. Due to small size of the sample the signal-to-noise ratio was rather small at these measurements. Thus, for comparison, the results of measurements performed in stationary field, up to 7 T, at 2 K for a non-oriented sample, are superimposed in Fig.8 onto the data obtained in pulse field. A good agreement between both data sets can be noticed. The pulse field measurements show much higher hysteresis of the transition field (4 - 4.5 T), which can be attributed to fast change of the field during experiment. The differential magnetic susceptibility $dM/dH$ is almost constant in the field range $H_{sr} < H < 11$ T, then decreases and reaches zero at 13.5 T. The saturation field for 4.2 K, $H_{sat} = 12.3$ T, was estimated as a middle of the field region in which $dM/dH$ decreases.

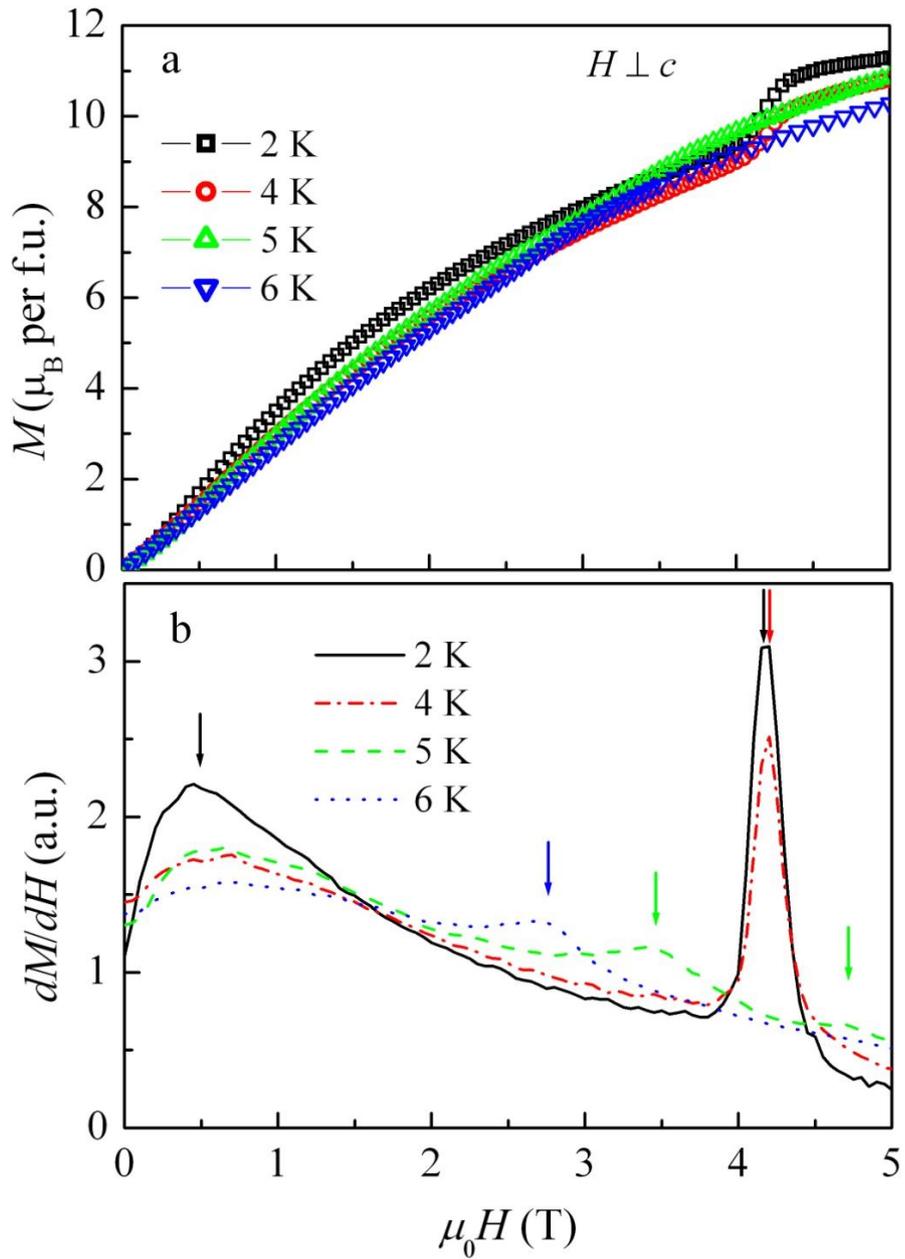

**Fig.6.** (a) Magnetic field dependencies of magnetization $M(H)$ of GdCr$_3$(BO$_3$)$_4$ measured at 2, 4, 5 and 6 K for $H\perp c$. (b) Derivatives $dM/dH(H)$ of these dependencies. Arrows indicate features on the $dM/dH(H)$ curves.

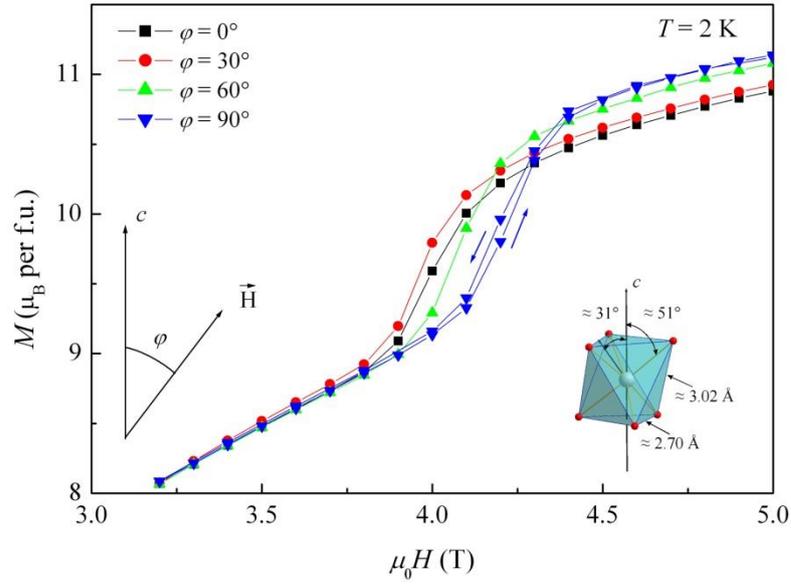

**Fig.7.** Magnetic field dependencies of magnetization $M(H)$ of GdCr$_3$(BO$_3$)$_4$ measured at 2 K in the vicinity of the spin-reorientation phase transition for different orientations of the external magnetic field with respect to the $c$ axis. An attempt to detect hysteresis of the spin-reorientation transition was performed for $\varphi = 90°$ only. The CrO$_6$ octahedron is depicted and the average O-O distances and specific angles with respect to the $c$ axis are given.

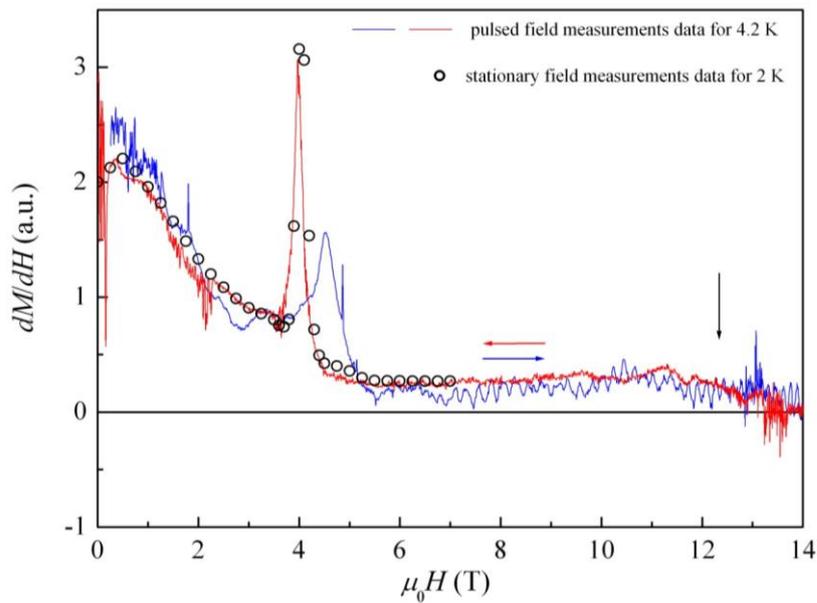

**Fig.8.** Differential magnetic susceptibility $dM/dH(H)$ measured in pulse magnetic field (blue and red lines) at 4.2 K and in stationary magnetic fields (symbols) at 2 K for a non-oriented GdCr$_3$(BO$_3$)$_4$ sample. Blue and red arrows indicate the data measured in increasing

and decreasing field, respectively. Black arrow indicates the saturation field.

### 3.5. Antiferromagnetic resonance

Figure 9 shows the frequency-field dependence of antiferromagnetic resonance (AFMR) spectrum of GdCr$_3$(BO$_3$)$_4$ obtained at 4.2 K in frequency range 17-142 GHz for $H\|c$ (red open circles) and $H\perp c$ (blue closed circles). Analysis of the low frequency part (17-38 GHz) of this dependence was reported recently [26]. It was found that AFMR spectrum can be described by the following expressions:

$$\left(\frac{\nu_\|}{\gamma}\right)^2 = \left(\frac{\Delta}{\gamma}\right)^2 + H^2, \quad \frac{\nu_\perp}{\gamma} = H, \tag{5}$$

derived in frames of a model of two-sublattice antiferromagnet with easy-plane anisotropy [65] (red and blue solid lines) with the following parameters: gap $\Delta = 25.5\pm0.5$ GHz, and gyromagnetic ratio $\gamma = 27.9\pm0.25$ GHz/T, which corresponds to the effective $g$-factor $g = 2.00\pm0.01$ ($\gamma = g\mu_B/h$). One can see a good agreement between the experimental data for $H < H_{sr}$ and the dependencies calculated by using the expressions (5).

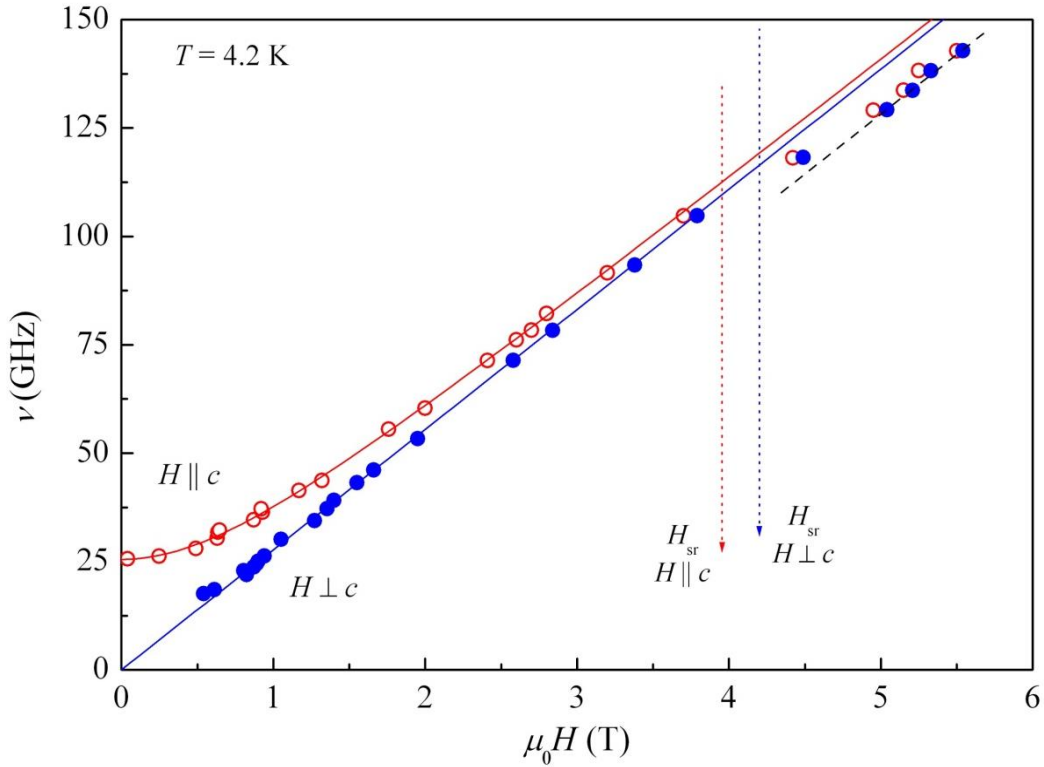

**Fig.9.** The frequency-field dependence of AFMR spectrum of GdCr$_3$(BO$_3$)$_4$ obtained at 4.2 K for $H\|c$ (red open circles) and $H\perp c$ (blue closed circles). Solid lines were calculated within the model of two-sublattice easy-plane antiferromagnet. Dotted arrows point out the field of the phase transition for both directions ($H\|c$ and $H\perp c$).

For $H$ larger than the spin-reorientation transition some deviation of the theoretical curves from the experimental data is seen. Nature of this deviation is not clear for now. No additional AFMR modes were detected in the whole frequency and field ranges. The value of the effective magnetic anisotropy field of the studied compound ($H_A \approx 0.05$ T) was estimated by using the following equation $\Delta/\gamma \approx (2H_E H_A)^{1/2}$ and taking the gap value given above and the exchange interaction field $2H_E \approx 17$ T. No evidence of additional magnetic anisotropy in the easy plane was detected.

*3.6. Heat capacity*

Figure 10 shows temperature dependence of the specific heat $C(T)$ of GdCr$_3$(BO$_3$)$_4$ measured in temperature range 2-300 K at zero magnetic field. Sharp λ-anomaly at $7.15 \pm 0.5$ K is attributed to the magnetic phase transition to the antiferromagnetic state. The fact that only one λ-anomaly is observed means that the sample has no admixtures of the monoclinic which were found for compound SmCr$_3$(BO$_3$)$_4$ [24]. It was assumed that $C(T)$ consists of lattice $C_{lat}(T)$ and magnetic $C_{mag}(T)$ contributions. The lattice contribution was estimated by fitting the expression (B.1) to the $C(T, H=0T)$ dependence, measured for 50 K < $T$ < 300 K (for details see Appendix B).

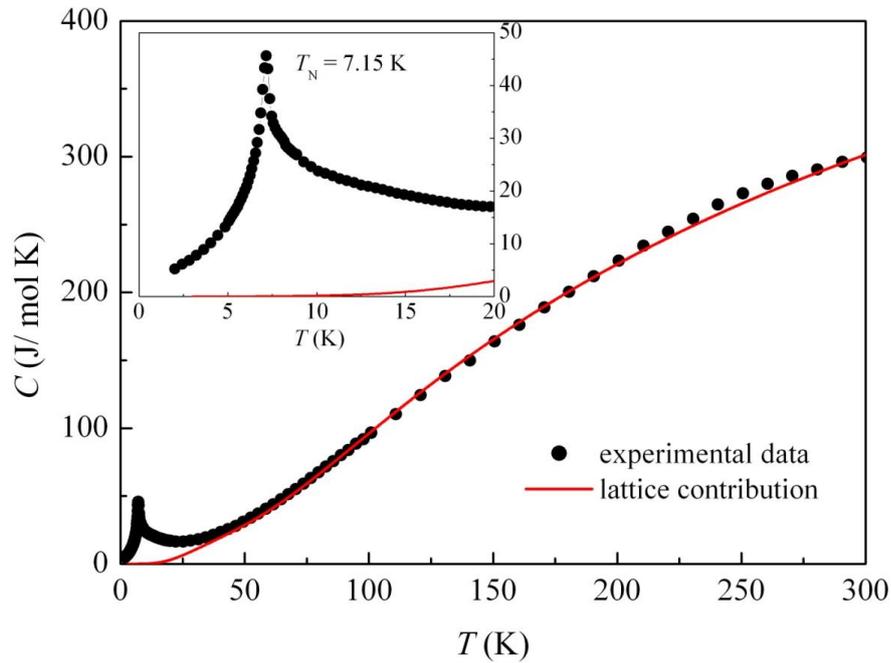

**Fig.10.** Temperature dependence of the specific heat $C(T)$ of GdCr$_3$(BO$_3$)$_4$ measured in zero magnetic field (circles). Solid line is the calculated lattice contribution (see Appendix B). The inset shows the low-temperature part of $C(T)$.

Magnetic heat capacity $C_{mag}(T)$ was determined by extracting the lattice contribution $C_{lat}(T)$ from the experimental data (Fig.11). A pronounced tail can be seen on $C_{mag}(T)$ well above $T_N$.

It is worth to compare this result with the data on the heat capacity of Heisenberg spin $S = 3/2$ chain obtained in Ref. [66], where dependencies $C(kT/J)$ for ferromagnetic and antiferromagnetic spin chains for several spin values up to 5/2 were calculated and tabulated. The red and blue dashed lines in Fig.11 present the specific heat for the ferromagnetic $J/k = -6.8$ K and antiferromagnetic $J/k = 4.3$ K chains, respectively. The heat capacity of AFM chain seems to describe better the experimental data. It should be noted that the results obtained in [66] for heat capacity of Heisenberg chains do not take into account any interchain exchange interactions in the Cr subsystem and any contribution of the Gd subsystem.

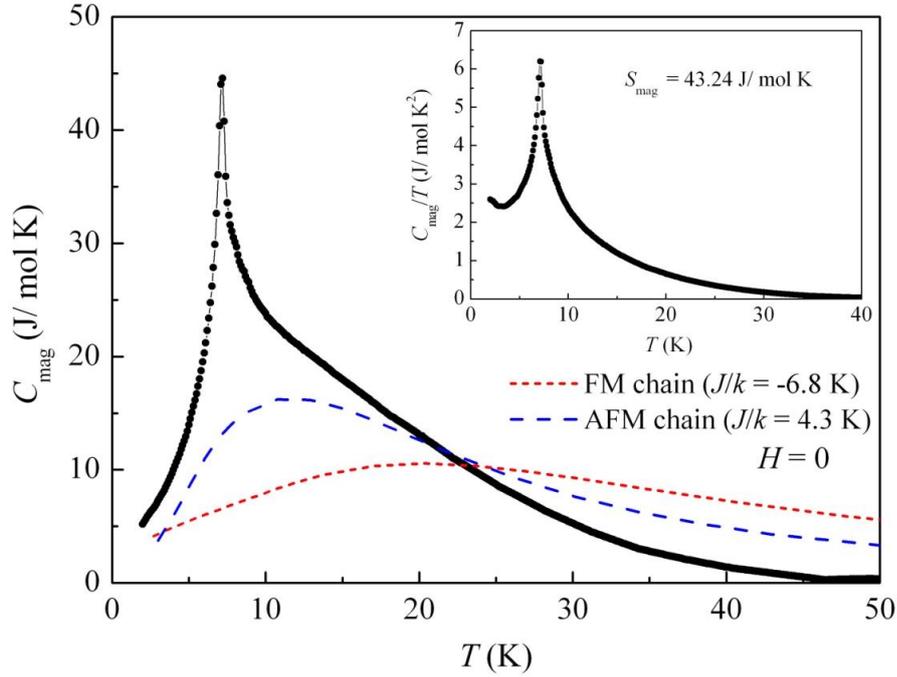

**Fig.11.** Temperature dependence of the magnetic part of the specific heat $C_{mag}(T)$ of GdCr$_3$(BO$_3$)$_4$ (black circles) at zero magnetic field. Dashed lines present the specific heat of ferromagnetic (red) and antiferromagnetic (blue) $S = 3/2$ Heisenberg chains with $J/k = -6.8$ K and $J/k = 4.3$ K, respectively. The inset shows the $C_{mag}(T)/T$ dependence for $H = 0$.

Moreover, for calculation of the lattice contribution, it was assumed that magnetic part of the heat capacity is negligible above 50 K. Hence, precise quantitative analysis of $C_{mag}(T)$ is hampered. Inset in figure 11 shows $C_{mag}/T$ vs $T$ dependence. The value of magnetic entropy $S_{mag} = 43.24$ J/(mol K) was evaluated by integrating the $C_{mag}(T)/T$ function. The expected value of magnetic entropy for GdCr$_3$(BO$_3$)$_4$ can be calculated independently by using the following expression,

$$S_{\text{mag}} = 3R\ln(2S_{Cr} + 1) + R\ln(2S_{Gd} + 1) = 51.87 \text{ J/(mol K)}, \qquad (6)$$

where $R = N_A k = 8.314$ J/(mol K) is the universal gas constant, $S_{Cr} = 3/2$ and $S_{Gd} = 7/2$ are spin values for Cr$^{3+}$ and Gd$^{3+}$ ions, respectively. Here the spin values instead of the total angular momentum values are used, because the orbital momentum of $Cr^{3+}$ is "quenched" by the ligand field and the orbital momentum of $Gd^{3+}$ is zero. The experimentally determined value of the magnetic entropy is less than the theoretical one by 17 % (8.61 J/(mol K)). Partly, this discrepancy can be explained by taking into account the fact that the $C_{mag}(T)$ for $T < 2$ K is unknown and it can give considerable contribution into magnetic entropy. Moreover, as inset to Fig.11 illustrates, upturn of the $C_{mag}(T)/T$ below 3 K is obvious. Such behavior can be a sign of either some phase transition below 2 K or a Schottky anomaly due to splitting of electronic levels of the gadolinium ions in the Cr-Gd exchange field.

The temperature dependencies of specific heat in nonzero external magnetic field were measured for two field orientations with respect to the $c$ axis of the crystal ($H\|c$ and $H\perp c$) and for several field values ranging from 0 to 9 T. No significant differences in the temperature dependencies of heat capacity for the two mentioned field orientations were found. The temperature dependencies of the magnetic specific heat in different magnetic field are shown in Fig.12. It can be seen that the magnetic ordering temperature demonstrates strong and nonmonotonic dependence on magnetic field, reaching a local minimum (4.45 K) in external magnetic field ≈ 4 T.

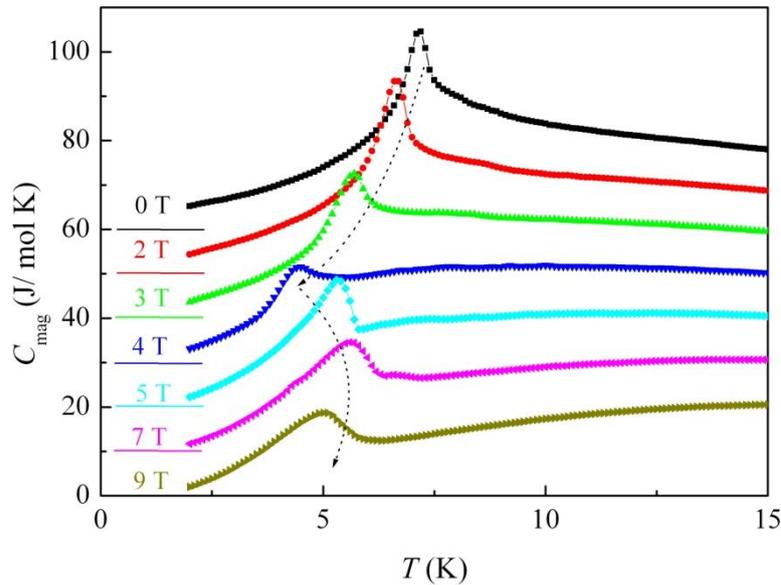

**Fig.12.** Temperature dependencies of the magnetic part of the specific heat $C_{mag}(T)$ of $GdCr_3(BO_3)_4$ obtained for a number of magnetic fields ($H\perp c$). The curves are equidistantly shifted along the ordinate axis by 10 J/(mol K) for clarity (the curve for zero field is shifted by 60 J/(mol K) and the curve for 9 T is not shifted at all). Horizontal lines show "0" heat capacity level for each curve. Dotted arrows show schematically changes of the magnetic ordering temperature with increase of the magnetic field.

*3.7. Electric polarization*

Figure 13 shows field dependencies of the longitudinal electric polarization of GdCr$_3$(BO$_3$)$_4$ measured along the *a* axis at 4.2 K in pulsed magnetic fields. The experimental data for different amplitudes of pulse are depicted by curves of different colour. The repeatability of the experiments is clearly seen. The magnetoelectric effect in the gadolinium chromium borates was found to be the same order of value as in the gadolinium iron borate [67]. Three features are visible on the Δ$P_a$($H_a$) curves. The first one is nonlinearity and hysteresis of polarization in low field region (below 0.5 T), which associated apparently with transformation of domain structure. The second one is change of a slope and small hysteresis of polarization in the vicinity of 1.4 T. The nature of this feature is not clear for now. The third feature has the form of a jump at 4.4 T and is associated with the spin-reorientation transition. In field ranges 1.4 - 4.0 T and 4.5 - 10 T, the polarization is nearly linear function of the magnetic field and reaches saturation in field close to 11 T. The nonzero value of polarization in the saturated paramagnetic state ($H > H_{sat} = 12.3$ T) indicates absence of an inversion center in the crystal and confirms the rhombohedral structure (*R*32). Moreover, absence of any noticeable decrease of the polarization during transition from ordered to saturated paramagnetic state points out that linear magnetoelectric effect in GdCr$_3$(BO$_3$)$_4$ is too small or even absent (linear magnetoelectric effect is forbidden in paramagnetic state). Alas, the magnetic point group for GdCr$_3$(BO$_3$)$_4$ is unknown, thus it is impossible to find out whether allowed the linear magnetoelectric effect or not in the ordered state. Meanwhile, we can say allowed the nonlinear magnetoelectric effect in the crystal or not knowing only point group, which is 32 in our case. There are ten nonzero components of the tensor characterizing the nonlinear magnetoelectric effect ($\beta_{ijk}$) with only four independent among them [68]. Therefore, electric polarization along the *a* axis of the crystal with point symmetry 32 is represented by the following equation (the linear magnetoelectric effect is omitted):

$$P_a = \tfrac{1}{2}\beta_{aaa}(H_a^2 - H_b^2) + \tfrac{1}{2}(\beta_{abc} + \beta_{acb})H_b H_c \,, \tag{7}$$

where indices *a*, *b* and *c* show components of *H* and *P* in respect to the crystal axes. Direction *b* is perpendicular to the *a* and *c* crystal axes simultaneously. In equation (7) components of *H* can be substituted by components of magnetization or antiferromagnetic vector [5,67,69]. Moreover, the contribution of each magnetic subsystem (Cr and Gd) should be written in the form of eq. (7) and total polarization is a sum of these contributions. It is impossible to determine four independent components of the tensor $\beta_{ijk}$ for both magnetic subsystems having data on polarization only for one geometry of the experiment.

Spontaneous polarization cannot be measured by means of used experimental technique. However, taking into account data for rare-earth iron borates [5,67], one can suggest that spontaneous polarization in the magnetically ordered state may also exist in $GdCr_3(BO_3)_4$.

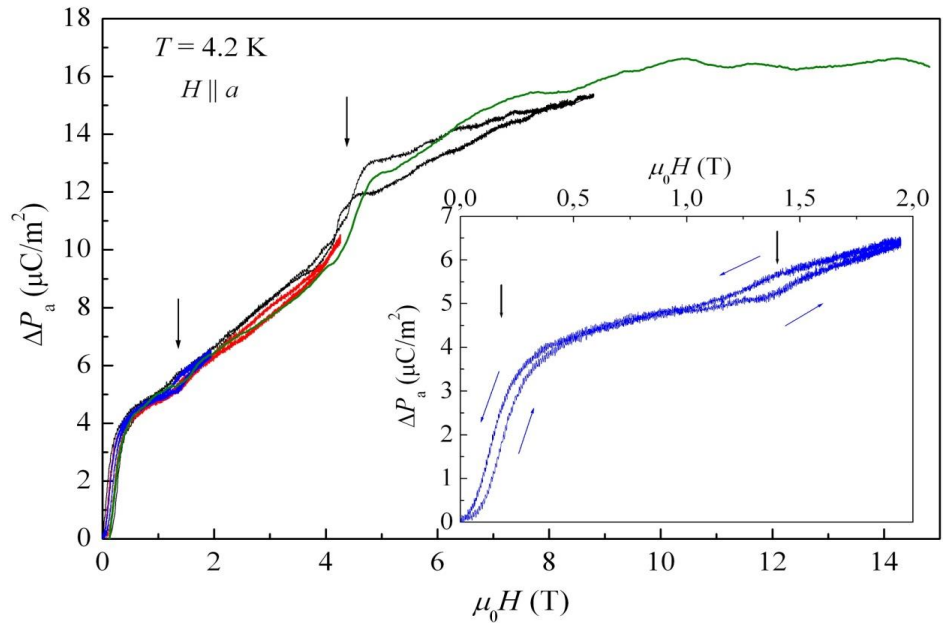

**Fig.13.** Magnetic field dependencies of the longitudinal polarization $\Delta P_a(H_a)$ of $GdCr_3(BO_3)_4$ measured along the *a* axis at 4.2 K in pulsed magnetic fields with different amplitudes of pulse. The inset shows one record for *H* up to 2 T. Vertical arrows indicate features on $\Delta P_a(H_a)$.

*3.8. Magnetic phase diagram*

Based on the experimental data described above, the magnetic phase diagram was constructed (Fig.14). It has already been discussed that the magnetic properties of the studied compound are almost isotropic and the presented phase diagram is valid for any direction of the external magnetic field with respect to the crystal axes. It can be seen that the temperature of magnetic ordering depends rather strongly and nonmonotonically on magnitude of the external magnetic field, reaching the local minimum (4.45 K) in the field ≈ 4 T. In magnetically ordered state, there are two magnetic phases that can be called low-field and high-field phases. We propose possible spin configurations for both magnetic phases (see Fig.14). It is suggested that the low-field phase is the antiferromagnetic phase with an EP anisotropy in which gradual rotation of the sublattices magnetizations of both subsystems happens when magnetic field increasing. In this case, no spin-reorientation transition with abrupt change of the magnetization is expected unless higher-order anisotropy is present, for example, cubic one which is characteristic of the octahedral environment. For studied compound, the CrO$_6$ octahedra are distorted (see inset in Fig.7), which, as we believe, makes some of directions close to the basal plane more favorable in comparison with the *c* axis. Existence of a high order anisotropy may give rise to the spin-reorientation transition not only for GdCr$_3$(BO$_3$)$_4$ but for whole subfamily of chromium borates *R*Cr$_3$(BO$_3$)$_4$ with rhombohedral structure, which is approved by features of isothermal magnetization for Nd and Eu chromium borates [21,22]. Both magnetic subsystems take part in the spin-reorientation transition due to the Cr-Gd exchange interaction and it is assumed that in the high-field phase the gadolinium subsystem is completely or almost completely magnetized. Further neutron diffraction measurements may define the structure of both phases precisely and help to understand the nature of spin-reorientation transition.

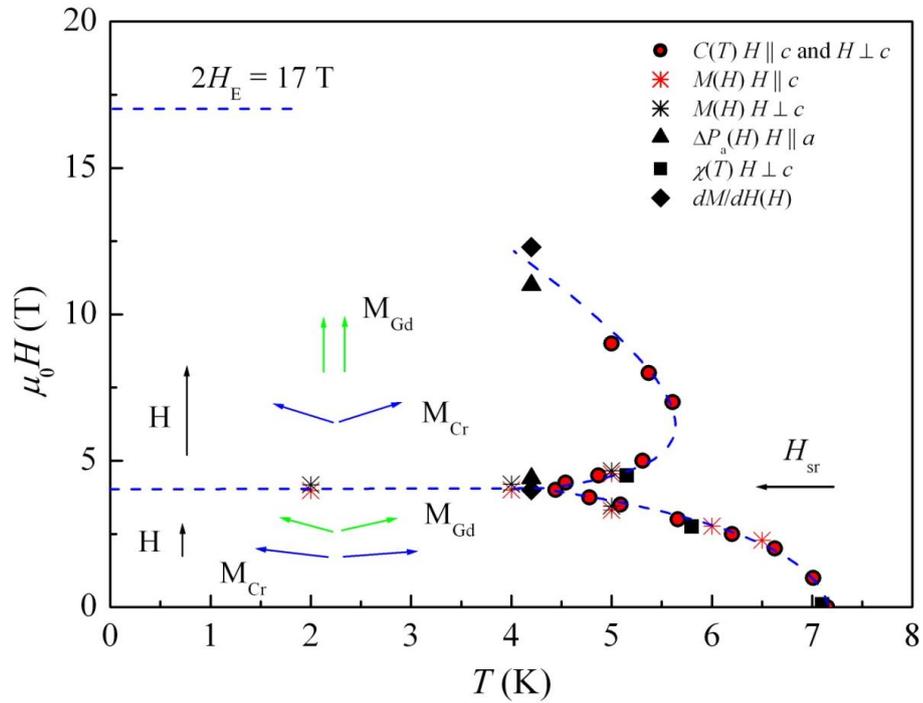

**Fig.14.** Magnetic phase diagram of GdCr$_3$(BO$_3$)$_4$. Phase boundaries are depicted by blue dashed curves (drawn by-eye lines). Spin configurations for low-field and high-field phases are proposed.

## 4. Conclusions

Studies of the magnetic, resonance, and magnetoelectric properties of GdCr$_3$(BO$_3$)$_4$ single crystal have been carried out. Survey of published data on the Cr$^{3+}$-Cr$^{3+}$ and Cr$^{3+}$-Gd$^{3+}$ exchange interaction and analysis of the magnetic susceptibility of GdCr$_3$(BO$_3$)$_4$ have been performed and, as a result, the magnitudes of the intrachain ($J/k$ = 4 K) and interchain ($J'/k \approx$ -0.7 K) interactions in the Cr subsystem have been determined and the magnitude of the exchange interaction between the subsystems ($J_{Cr-Gd}/k \sim$ -0.2 K) has been estimated. More precise value of the Neel temperature $T_N$ = 7.15 K has been obtained by using the results of heat capacity measurements. The values of the exchange field ($2H_E \approx$ 17 T) and the effective magnetic anisotropy field ($H_A \approx$ 0.05 T) of the studied sample have been estimated. Electric polarization $\Delta P_a$ in the longitudinal geometry of the experiment has been found. The spin-reorientation phase transition in the magnetically ordered state has been found. This transition exists for all directions in the crystal and transition field (3.9 – 4.2 T) weakly depends on the direction of the external magnetic field. Existence of a spin-reorientation phase transition is supposed to be due to the anisotropy of the $Cr^{3+}$ in distorted octahedral ligand field. Magnetic phase diagram has been constructed and spin configurations for the low-field and high-field phases have been proposed.

**Acknowledgments**
This work was partially supported by the National Science Centre, Poland, under project No. 2018/31/B/ST3/03289.

## Appendix A

Table 1. Structural and magnetic data for some oxides containing trivalent chromium ions located in sharing edges oxygen octahedral.

| Compound | $d_{Cr-Cr}$ (Å) | $d_{Cr-O}$ (Å) | ∠Cr-O-Cr (deg) | $J/k$ (K) | $T_N$ (K) | $\theta_{CW}$ (K) | $\mu_{eff}$ ($\mu_B$ per Cr) | Refs. |
|---|---|---|---|---|---|---|---|---|
| PbCrBO$_4$ | 2.97 | 1.94; 2.031 | 99.8; 94.0 | 4 | 8 | -45 | 3.7 | [38,39] |
| ZnCr$_2$O$_4$ | 2.9450 | 2.0825 | 90 | 36.9 | 12.5 | -390 | - | [40,41] |
| CdCr$_2$O$_4$ | 3.0289 | 2.1418 | 90 | 21.7 (15) | 7.8 | -71 | - | |
| LiCrSi$_2$O$_6$ | 3.064 | 1.992 | 98.3 | 4 | 11 | -28.7 | 3.7 | [42-45] |
| NaCrSi$_2$O$_6$ | 3.086 | 1.993 | 99.6 | 0.4 | 3 | -0.3 | 3.7 | |
| LiCrGe$_2$O$_6$ | 3.098 | 1.995 | 100.2; 98.7 | 0.6 | 3.7 | -5.7 | 3.7 | |
| NaCrGe$_2$O$_6$ | 3.140 | 2.004 | 101.2 | -2.6 | 6($T_C$) | +13 | 3.7 | |
| LiCrO$_2$ | 2.900 | - | 94.2 | 39 | 62 | -700 | - | [46-48] |
| PdCrO$_2$ | 2.923 | - | - | 23 | 108 | -500 | - | |
| NaCrO$_2$ | 2.970 | - | 95.3 | 20 | 46 | -290 | - | |
| CuCrO$_2$ | 2.975 | - | 96.6 | 11.4 | 27 | -199 | 3.88 | |
| AgCrO$_2$ | 2.984 | - | 96.6 | 9 | 34 | -168 | 3.94 | |
| KCrO$_2$ | 3.020 | - | 96.8 | 11.5 | 4.2 | -160 | - | |
| LiCr(MoO$_4$)$_2$ | 2.987 | 1.939; 1.969 | 99.7 | 6.3 | - | -79 | - | [49] |
| Cr$_2$Te$_4$O$_{11}$ | 3.01 | 1.99 | 102 | 4.3 | - | - | - | [50-52] |
| Cr$_2$TeO$_6$ | 2.9840 | 1.934; 1.908 | 98.68 | 33.2 | 93 | - | 3.4 | |
| Cr$_2$WO$_6$ | 2.9360 | 1.964; 1.897 | 97.98 | 44.5 | 45 | - | 3.6 | |
| CrVO$_4$ | 2.9890 | - | 97.7 | 22, 25 | 50 | -214 | - | [53-55] |
| CrPO$_4$ | 3.0290 | - | 97.9 | 20 | 23 | -80 | 3.9 | |
| NdCr$_3$(BO$_3$)$_4$ | 3.091 | 1.985; 1.996 | 101.46; 102.27 | 4 | 8 | -26 | 3.87 | [21-25, 37, this work] |
| SmCr$_3$(BO$_3$)$_4$ | 3.0856 | 1.982; 1.994 | 101.38; 102.18 | 2.8 | 7.8 | - | - | |
| EuCr$_3$(BO$_3$)$_4$ | 3.077 | 1.984; 1.995 | 101.46; 102.26 | 4.6 | 9.8 | -21.9 | 3.99 | |
| TbCr$_3$(BO$_3$)$_4$ | 3.0825 | 1.98; 1.991 | 101.46; 102.26 | 3.2 | 8.8 | -8 | 3,87 | |
| GdCr$_3$(BO$_3$)$_4$ | 3.0832 | 1.98; 1.992 | 101.43; 102.24 | (-6.8) 4.3 | 7.15 | +7 | 3,87 | |
| Cr$_2$O$_3$ | 2.89 | 2.02; 1.97 | 93.1 | 19 | 306 | -430 | 3.8 | [56] |

**Appendix B**

The lattice heat capacity was estimated by using expression (B.1), under assumption that above 50 K, the heat capacity is determined exclusively by the lattice contribution. The lattice heat capacity is a sum of contribution of acoustic and optical phonons. Initial values of energies of particular optical phonon branches, $\theta_i$ and number of vibration modes for each branch, $n_i$ were obtained by using data on IR and Raman spectroscopy for compounds belonging to this family [18,19,70-72]. The best fit of experimental data in the range 50-300 K was obtained for the parameters listed in the table B.1.

$$C_{lat} = kN_A \left[ 3n_D \left(\frac{T}{\theta_D}\right)^3 \int_0^{\frac{\theta_D}{T}} \frac{x^4 e^x}{(e^x-1)^2} dx + \sum_{i=1}^{7} n_i \left(\frac{\theta_i}{T}\right)^2 \frac{e^{\frac{\theta_i}{T}}}{\left(e^{\frac{\theta_i}{T}}-1\right)^2} \right], \quad (B.1)$$

where $k$ – Boltzmann constant, $N_A$ - Avogadro constant, $\theta_D$ and $\theta_i$ - Debye and Einstein temperatures respectively, $n_D$ and $n_i$ number of corresponding vibration modes.

**Table B.1.** The parameters $\theta_D$, $n_D$, $\theta_i$, $n_i$ that were used for calculating the lattice heat capacity of $GdCr_3(BO_3)_4$.

| $i$ | Debye | 1 | 2 | 3 | 4 | 5 | 6 | 7 |
|---|---|---|---|---|---|---|---|---|
| $n_D$, $n_i$ | 3 | 3 | 21 | 9 | 8 | 4 | 4 | 8 |
| $\theta_D$, $\theta_i$ ; K | 192 | 128 | 411 | 681 | 996 | 1088 | 1395 | 1843 |